\documentclass[prb,twocolumn,groupedaddress,showpacs]{revtex4-1}
\usepackage{graphicx}
\usepackage{amsmath}

\begin{document}


\title{$^{75}$As NMR study of overdoped CeFeAsO$_{0.8}$F$_{0.2}$}


\author{Damian Rybicki$^{1,2}$} 
\email[]{rybicki@physik.uni-leipzig.de} 
\author{Thomas Meissner$^1$}
\author{Grant V.M. Williams$^3$}
\author{Shen Chong$^4$}
\author{Marc Lux$^1$}
\author{J\"urgen Haase$^1$}
\affiliation{$^1$ Faculty of Physics and Earth Science, University of Leipzig, Linn$\acute{e}$stra\ss e 5, 04103 Leipzig, Germany}
\affiliation{$^2$ Faculty of Physics and Applied Computer Science, AGH University of Science and Technology, 30-059 Krak\'{o}w, Poland}
\affiliation{$^3$ SCPS, Victoria University of Wellington, PO Box 600, Wellington 6012, New Zealand}
\affiliation{$^4$ Industrial Research Ltd., PO Box 31310, Lower Hutt 5040, New Zealand}

\date{\today}

\begin{abstract}
We report the results from a $^{75}$As NMR study on the overdoped iron pnictide superconductor, CeFeAsO$_{0.8}$F$_{0.2}$. Two peaks are observed in the NMR spectra for temperatures as high as 100 K, which is above the superconducting transition temperature of 39 K and hence they cannot be attributed to the effect of vortices in the superconducting state as previously suggested [A. Ghoshray, et al., Phys. Rev. B 79, 144512 (2009)]. The temperature dependence of the $^{75}$As NMR shifts for the two peaks is consistent with hyperfine coupling from magnetic Ce to As and with different Curie-Weiss temperatures. The appearance of two Curie-Weiss temperatures where one is near zero and the other is $\sim$-18 K suggests that the two peaks arise from ordered and disordered regions where the disorder could be due to regions with significant fluorine site disorder and/or clusters of oxygen vacancies. Similar to some other studies on ReFeAsO$_{1-x}$F$_x$ (Re is a rare earth) we find that there are more than one $^{75}$As spin-lattice and spin-spin relaxation rates that can be attributed to ordered and disordered regions in the ReO layer and suggests that inhomogeneity is a common feature in the ReFeAsO$_{1-x}$F$_x$ superconductors.
\end{abstract}


\maketitle


Iron pnictides are a new family of high temperature superconductors (HTSC) with the second highest superconducting transition temperatures after the cuprates. They have attracted a considerable amount of attention because of the complex Fermi surface and similarities to the cuprates (e.g. the parent compounds display magnetic order). This raises the possibility that superconductivity is mediated by antiferromagnetic spin fluctuations. The CeFeAsO$_{1-x}$F$_x$ compounds belong to the 1111 iron oxypnictide family and they are similar to other pnictides where they contain FeAs layers (with one As lattice site) separated by a CeO layer. In the parent compound Fe orders antiferromagnetically (AF) at about 140 K.\cite{Zhao2008} Upon F doping (electron doping) the AF transition temperature decreases and vanishes completely as the F content increases up to about 6\% where superconductivity emerges with the highest $T_{\rm c}$ of about 40 K.\cite{Zhao2008, Shiroka2011} Superconductivity has been observed in samples with F concentrations of up to $\sim$20\%, however the superconducting dome is flat compared to the cuprates.\cite{Zhao2008} The Ce ions can order magnetically at low temperatures in contrast to La in LaFeAsO$_{1-x}$F$_x$, which was the first discovered superconducting system in the 1111 family.\cite{Kamihara2008} This offers the possibility to study the relation between magnetism of the rare earth and superconductivity.  

Most of the current research is focused on the lower doping region especially close to the AF to superconducting phase boundary with the aim to better understand the nature of this transition and to see whether these phases coexist. The overdoped region gets less attention, which is due to the fact that it is more difficult to make clean samples with higher F content.\cite{Chong2008} In several iron pnictides, particularly at the higher doping levels, $^{75}$As nuclear magnetic resonance (NMR) experiments showed that at a certain temperature the signal comes from more than one As environment. This was manifested by more than one $^{75}$As line in the NMR spectrum or two different spin-lattice relaxation times $T_1$.\cite{Ghoshray2009, Lang2010, Kobayashi2010, Mukuda2010, Yamashita2010, Baek2011} Explaining such effects can be crucial for understanding the nature of superconductivity in iron pnictides. There is clearly a need for more $^{75}$As NMR measurements on aligned samples to better understand the electronic and magnetic properties of CeFeAsO$_{1-x}$F$_{x}$.

In this article we present the results from temperature dependent $^{75}$As NMR shift, spin-lattice and spin-spin relaxation rate measurements on aligned CeFeAsO$_{0.8}$F$_{0.2}$. The measurements were made on aligned powder samples because there are no sizeable single crystals of this family of iron pnictides. However, the data quality is improved compared to a previous report \cite{Ghoshray2009} most likely due to better sample alignment. Therefore, we were able to carry out measurements with the external magnetic field $B_0$ parallel and perpendicular to the $c$ axis. Below ~100 K the $^{75}$As central line splits into two and we provide a possible explanation for this behavior.   

\section{Experimental}
Polycrystalline CeFeAsO$_{1-x}$F$_x$ oxypnictide was prepared by reacting stoichiometric amounts of presynthesized CeAs and FeAs with Fe, As, CeO$_2$ or Fe$_2$O$_3$, and CeF$_3$ in evacuated and sealed quartz tubes. The cerium oxypnictide components were thoroughly ground and pressed into a pellet prior to heat treatment. The pellet was first reacted at 1000 $^{\circ}$C for 24 hours, followed by an intermediate grinding and repressing into a pellet, before a final reaction at 1180 $^{\circ}$C for 50 hours. X-ray diffraction (XRD) of the as-synthesized overdoped CeFeAsO$_{0.8}$F$_{0.2}$ sample showed the presence of Ce$_3$Fe$_4$O$_3$ and traceable amounts of FeAs, FeAs$_2$ impurities. From XRD data we estimated that sample contained 8\% of Ce$_3$Fe$_4$O$_3$ and about 3\% of FeAs and FeAs$_2$. Alignment of the crystallite powder of CeFeAsO$_{0.8}$F$_{0.2}$ was carried out by mixing 100 mg of a finely ground oxypnictide powder with an epoxy and placing it in a 1 T magnet. The sample was rotated at a speed of 60 rpm with the magnetic field applied perpendicular to the rotating axis. XRD rocking curve analysis revealed a (004) full width half maximum of 7.0$^{\circ}$ using Co K$\alpha$ radiation. The superconducting transition temperature was measured using a SQUID magnetometer and found to be 39 K.

Most of $^{75}$As (nuclear spin, $I$=3/2) NMR measurements were carried out in a magnetic field of $B_{0}$=11.75 T. Additionally, we measured $^{75}$As NMR spectra at 7.07 T and 17.62 T in order to know the field dependence of the line shape. NMR shifts $^{75}K$, spin-lattice $T_{1}$ and spin-spin $T_{2}$ relaxation times were measured at temperatures ranging from 297 K to 10 K for different orientations of the aligned sample with respect to the external magnetic field, $B_{0}$ i.e. for c$\parallel$$B_{0}$ and c$\perp$$B_{0}$  (c axis parallel/perpendicular to $B_{0}$). A frequency stepped spin echo technique was used for recording the spectra with the pulse sequence $\tau_{\pi/2}-\tau-\tau_{\pi}$, where the echoes were integrated and plotted versus the frequency. 

The spin-spin relaxation time, $T_{2}$, was measured by varying the pulse spacing, $\tau$. The echo decay had an exponential form for both field orientations that can be written as, 
\begin{equation} 
M(\tau) = M(0)\cdot exp\left(\frac{-2\tau}{T_2}\right).
\label{eq:t2}
\end{equation}

The spin-lattice relaxation time, $T_1$, was measured using the inversion recovery method. The magnetization recovery signal could be fitted with the theoretical formula for the central transition of the $I$=3/2 nucleus,
\begin{equation} 
M(t) = M(\infty) \left[1-f\left(0.1\cdot exp\left(\frac{-t}{T_1}\right)+0.9\cdot exp\left(\frac{-6t}{T_1} \right) \right) \right],
\label{eq:t1}
\end{equation}
where $t$ is the time between the inversion pulse and the spin echo sequence, and $f$ is 2 for perfect inversion. 

The NMR frequency of the central transition line for $^{75}$As in a high magnetic field with the quadrupolar interaction treated as a perturbation to  second order and for the axial symmetry of the electric field gradient (EFG), i.e. asymmetry parameter $\eta$ = 0, can be written as,
\begin{equation} 
\label{eq:freq}
\nu = ^{75}\gamma B_0 \left( 1 + K \right)  + \Delta  \nu_q ^{ \left( 2 \right)} \left( \theta, B_0 \right), 
\end{equation}
where $^{75}\gamma$ is the gyromagnetic ratio of $^{75}$As, $K$ is the shift, $\Delta \nu_q ^{ \left( 2 \right)}$ is the second order quadrupolar contribution, which is a function of the external field and angle, $\theta$, between the external field and $c$ axis. With the field parallel and perpendicular to the $c$ axis (c$\parallel$$B_{0}$ and c$\perp$$B_{0}$, respectively) Eq. \ref{eq:freq} gives,
\begin{equation} 
\label{eq:Kparall}
\nu_{\parallel} = ^{75}\gamma B_0 \left( 1 + K_{\parallel} \right)
\end{equation}
\begin{equation}
\label{eq:Kperp}
\nu_{\perp} = ^{75}\gamma B_0 \left( 1 + K_{\perp} \right) + \frac{3 \cdot \nu_{q}^{2}}{16 \cdot ^{75}\gamma B_0},
\end{equation} 
where $\nu_{q}$ is the quadrupolar frequency measured for the c$\parallel$$B_{0}$ orientation.
The shift $^{75}K$ can be calculated from Eq. \ref{eq:Kparall} and \ref{eq:Kperp} and by measuring the frequency ($^{75}\gamma B_0$) of the reference compound. The shifts were referenced to NaAsF$_6$ using the procedure described by Harris et al.\cite{Harris2008}

In general the total NMR shift (after subtracting the quadrupolar contribution) is given by the formula,
\begin{equation}
K_{\parallel, \perp}(T) = K_{L,\parallel, \perp} + K_{s,\parallel, \perp}(T) + K_{M,\parallel, \perp},
\label{eq:shift}
\end{equation}
where $K_{L}$ is the orbital shift, $K_{s}$ is the spin (Knight) shift, which can be written as $K_{s}=A_{\parallel, \perp}\chi_{s}$ where $A$ is the hyperfine coupling constant and $\chi_{s}$ is the uniform electronic spin susceptibility. $K_{M}$ is the Meissner (diamagnetic) shift which appears only below $T_c$ and it is small when compared to the very large low temperature shifts that we observe in our sample. Therefore, we will restrict our discussion to $K_{s}$ and $K_{L}$. All shifts are orientation dependent owing to a directional dependent hyperfine coupling constants. $K_{L}$ is expected to be temperature independent while $K_{s}$ can change significantly with temperature and reflects the temperature dependence of $\chi_{s}$. $K_{L}$ and $K_{s}$ are of magnetic origin and field independent.

\section{Results and discussion}
\subsection{Temperature and field dependence of $^{75}$As spectra}
\begin{figure}
\includegraphics[scale=0.75]{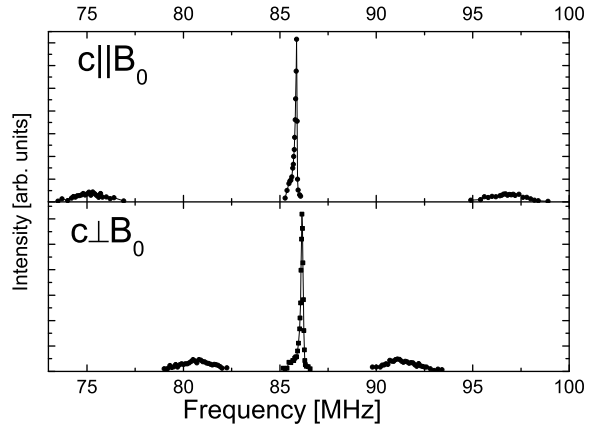}
\caption{$^{75}$As spectra of CeFeAsO$_{0.8}$F$_{0.2}$ at 11.75 T and at 297 K for two orientations of the external field with respect to the $c$ axis,  c$\parallel$$B_{0}$ and c$\perp$$B_{0}$.\label{fig:spectra}}
\end{figure}
Fig. \ref{fig:spectra} shows a typical frequency stepped $^{75}$As spectrum of the aligned sample for c$\parallel$$B_{0}$ and c$\perp$$B_{0}$ measured at room temperature at 11.75 T. One can see a sharp, central line due to the central transition $m=+1/2$ to $-1/2$ and satellite lines arising from the $\pm1/2$ to $\pm3/2$ transitions. The satellite lines are broad, which is also observed in the $^{63}$Cu NMR spectra in the superconducting cuprates\cite{Rybicki2009}. From c$\parallel$$B_{0}$ we calculate the quadrupole splitting, $\nu_q$ which amounts to 10.9$\pm$0.2 MHz, and within the measurement precision it does not change as the temperature is reduced to 10 K. Half of the distance between the satellite lines for c$\perp$$B_{0}$ amounts to 5.35$\pm$0.05 MHz which is expected for an axially symmetric EFG and it is consistent with the tetragonal crystal structure\cite{Zhao2008}. The value of $\nu_q$ is in agreement with that obtained for CeFeAsO$_{0.86}$F$_{0.16}$ \cite{Ghoshray2009} and similar to values obtained for LaFeAsO$_{1-x}$F$_{x}$ \cite{Grafe2008, Nakai2009} and CeFeAs$_{0.95}$P$_{0.05}$O.\cite{Sarkar2012} The second order quadrupole contribution, calculated for 11.75 T amounts to 0.26 MHz and it is also temperature independent.
\begin{figure}
\includegraphics[scale=0.75]{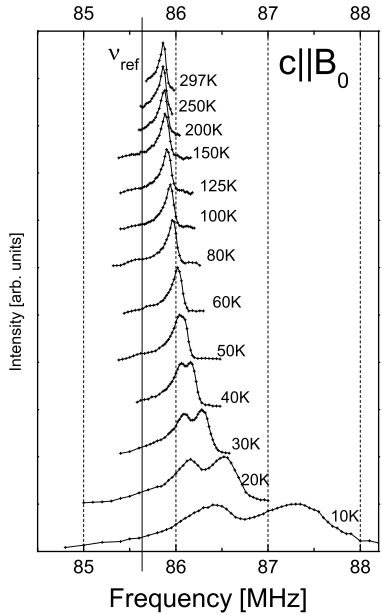}
\caption{Temperature dependence of $^{75}$As spectra of CeFeAsO$_{0.8}$F$_{0.2}$ for c$\parallel$$B_{0}$ measured at 11.75 T where $\nu_{ref}$ is the reference frequency. \label{fig:Tparallel}}
\end{figure}

\begin{figure}
\includegraphics[scale=0.75]{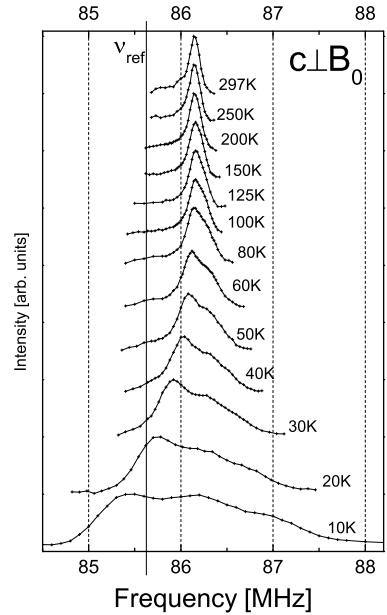}
\caption{Temperature dependence of $^{75}$As spectra of CeFeAsO$_{0.8}$F$_{0.2}$ for c$\perp$$B_{0}$ measured at 11.75 T where $\nu_{ref}$ is the reference frequency.\label{fig:Tperp}}
\end{figure}

\begin{figure}
\includegraphics[scale=0.75]{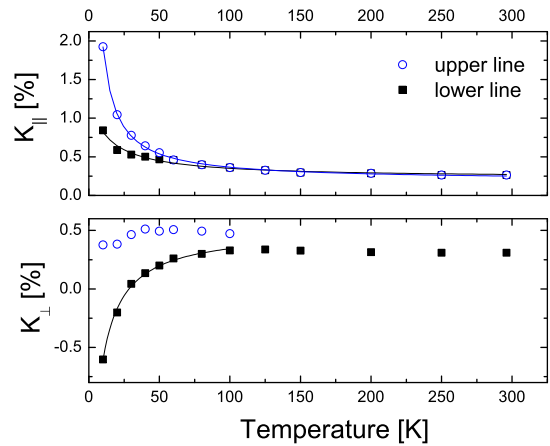}
\caption{Temperature dependence of shifts, $K_{\parallel}$ and $K_{\perp}$ (after subtraction of the second order quadrupolar contribution, see text). Lower/upper line refers to the lower/higher frequency line (see Figs. \ref{fig:Tparallel} and \ref{fig:Tperp}). Solid lines are Curie-Weiss fits to the data (Eq. \ref{CW}). 
\label{fig:shifts}}
\end{figure}

Now we turn to the temperature dependence of the $^{75}$As spectra and shifts for c$\parallel$$B_{0}$ and c$\perp$$B_{0}$ orientations presented in Fig. \ref{fig:Tparallel} and Fig. \ref{fig:Tperp}, respectively. For both orientations at higher temperatures the shift, $^{75}K$ (Fig. \ref{fig:shifts}), changes very slowly and then around 100 K starts to change much faster with no noticeable features around $T_{c}$. A similar temperature dependence was observed from $^{19}$F NMR measurements on SmFeAsO$_{0.85}$F$_{0.15}$ \cite{Prando2010}, and $^{75}$As NMR measurements on NdFeAsO$_{0.85}$F$_{0.15}$ \cite{Jeglic2009}. On the other hand the temperature dependence of $^{75}K$ for LaFeAsO$_{1-x}$F$_{x}$ is very different\cite{Grafe2008, Imai2008}, where it changes much less with temperature and is similar to the AeFe$_2$As$_2$ iron pnictides, where Ae is an alkaline earth (122 family). This must be due to the fact that Ce and Nd ions have a magnetic moment coming from $4f$ electrons \cite{Zhao2008, Chen2008, Qiu2008, Lynn2009} that affects the NMR shift while La is not magnetic \cite{delacruz2008}. Comparing the total temperature change of $^{75}K$ one notices that it is bigger for the Nd 1111 oxypnictide \cite{Jeglic2009} than for the Ce compound (Fig. \ref{fig:shifts}). This correlates with the free ion effective magnetic moment in the paramagnetic state, which is 3.62$\mu_B$ for Nd$^{3+}$ and  2.54$\mu_B$ for Ce$^{3+}$. For samples with small F doping it has been found that the rare earth orders magnetically at low temperatures (few K) with the ordered moment $\mu_{Nd}$=1.55$\mu_B$ \cite{Qiu2008} and $\mu_{Ce}$=0.94$\mu_B$ \cite{Zhao2008}.  

With decreasing temperature there is also very significant line broadening and a second line appears. Interestingly, splitting into two lines becomes visible at different temperatures for different field orientations. The resultant NMR shifts for both peaks and orientations are plotted in Fig. \ref{fig:shifts} where it can be seen that there is Curie-Weiss-like behavior.

Since there is only one As lattice site and no magnetic ordering is expected at high temperatures one expects only one central line in the $^{75}$As spectrum. However, there have been several studies so far that reported more than one line. Two lines were reported by Ghoshray et al. \cite{Ghoshray2009} for optimally doped CeFeAsO$_{0.84}$F$_{0.16}$, but the second line appeared only below $T_c$ and the lines were attributed to nuclei in superconducting and normal regions of the sample. A $\mu$SR study on underdoped CeFeAsO$_{1-x}$F$_{x}$ reported the coexistence of magnetic and superconducting states at low temperatures \cite{Sanna2010}. Nuclear quadrupole resonance (NQR) studies of LaFeAsO$_{1-x}$F$_{x}$ and SmFeAsO$_{1-x}$F$_{x}$ reported two lines in the underdoped \cite{Lang2010} and overdoped region \cite{Kobayashi2010}. It has also been shown that two different $T_{1}$ values are necessary to fit the LaFeAsO data \cite{Mukuda2009, Yamashita2010a}. Also, in the Ca(Fe$_{1-x}$Co$_x$)$_2$As$_2$ system an $^{75}$As NQR study showed that for optimally doped and overdoped samples a shoulder in the NQR line appears at relatively high temperatures indicating the existence of two As sites with different local environments.\cite{Baek2011} It should be noted that there are very few NMR studies of 1111 oxypnictides in the overdoped region making it difficult to fully compare our results with other systems.

It is likely that the Curie-Weiss behavior in the $^{75}$As data is due to hyperfine coupling of the Ce moments to $^{75}$As. Then Eq. \ref{eq:shift}  can be rewritten as\cite{Jeglic2009},
\begin{equation}
K_{\parallel, \perp}(T) = K_{L,\parallel, \perp} + \frac{A_{\parallel, \perp}^{FeAs} \chi_{FeAs}(T)}{N_A \mu_B} + \frac{A_{\parallel, \perp}^{Ce}, \chi_{Ce}(T)}{N_A \mu_B}
\label{eq:shiftCW}
\end{equation}
where the two last terms constitute the spin (Knight) shift $K_s$, $N_A$ is the Avogadro number, $\mu_B$ is the Bohr magneton, $\chi_{FeAs}$ and $\chi_{Ce}$ are the susceptibilities of itinerant electrons in the FeAs layer and of localized Ce moments, respectively. We do not know the temperature dependence of $\chi_{FeAs}$, however, for La 1111 oxypnictides, which do not have magnetic rare earth ions, the total temperature dependence of the measured shift $K_{s, \perp}$ is $\approx$15 times smaller than for our sample \cite{Imai2008, Grafe2008}. Therefore, since $\chi_{Ce}$ has a Curie-Weiss temperature dependence $K_{\parallel, \perp}$ can be written as,
\begin{equation}
\label{CW}
K_{\parallel, \perp}=b_{\parallel, \perp}/(T- \Theta) + a_{\parallel, \perp}, 
\end{equation}
which we use to fit the data. Fig. 4 shows such fits down to the lowest temperatures since we do not see any change in the behavior of $^{75}K$ around $T_{c}$. We did not fit the $K_{\perp}$ data for the upper line because the experimental uncertainty was too large. For $K_{\parallel}$ we fitted both curves simultaneously while maintaining a common $b_{\parallel}$, which is expected since the Ce effective moment should be the same for both peaks. 

For c$\parallel$$B_{0}$ we find that \mbox{$\Theta$=-18.4$\pm$1.6} K and $\Theta$=0.17$\pm$0.44 K for the lower and upper lines, respectively. Our Curie-Weiss temperature obtained for the lower peak is similar to $\Theta$=-13 K found from $^{75}K$ for NdFeAsO$_{0.85}$F$_{0.15}$ \cite{Jeglic2009} and $\Theta$=-11 K which was obtained from $^{19}F$ shifts for SmFeAsO$_{0.85}$F$_{0.15}$ \cite{Prando2010}. A Curie-Weiss-like behavior of $^{75}K$ was also found for NdFeAsO$_{0.6}$, but no values of $\Theta$ were given \cite{Yamashita2010}. The values for $a_{\parallel}$ are similar and they are 0.208$\pm$0.009 \% and 0.196$\pm$0.011 \% for the upper and lower lines, respectively, and $b_{\parallel}$ is 17$\pm$0.8 \%/K. The fit for $K_{\perp}$ is not as good as those for $K_{\parallel}$ and for the lower line we find that $\Theta$=-0.92$\pm$1.6 K, $a_{\perp}$=0.40$\pm$0.02 \%, and $b_{\perp}$=-11.1$\pm$1.7 \%/K. From the $b$ values we deduce that $A_{\parallel}/A_{\perp}$=-1.53$\pm$0.25 and the hyperfine coupling constant changes sign when the applied magnetic field orientation changes from parallel to perpendicular. The origin of the two lines is discussed below in section \ref{Origin}

Further understanding of the two lines as well as the line broadening mechanism can be obtained from measurements of the $^{75}$As spectrum in a broad field range ($B_{0}$=7.07 T, 11.75 T, 17.62 T), at different temperatures (40 K, 100 K and 297 K), and for both field orientations. We want to stress that compared to a report on optimally doped CeFeAsO$_{0.84}$F$_{0.16}$\cite{Ghoshray2009} for c$\perp$$B_{0}$ at 7 T we have smaller linewidths, which is most likely due to better alignment. We also did not observe any unexpected intensity changes in the normal state. 
\begin{figure}
\includegraphics[scale=0.75]{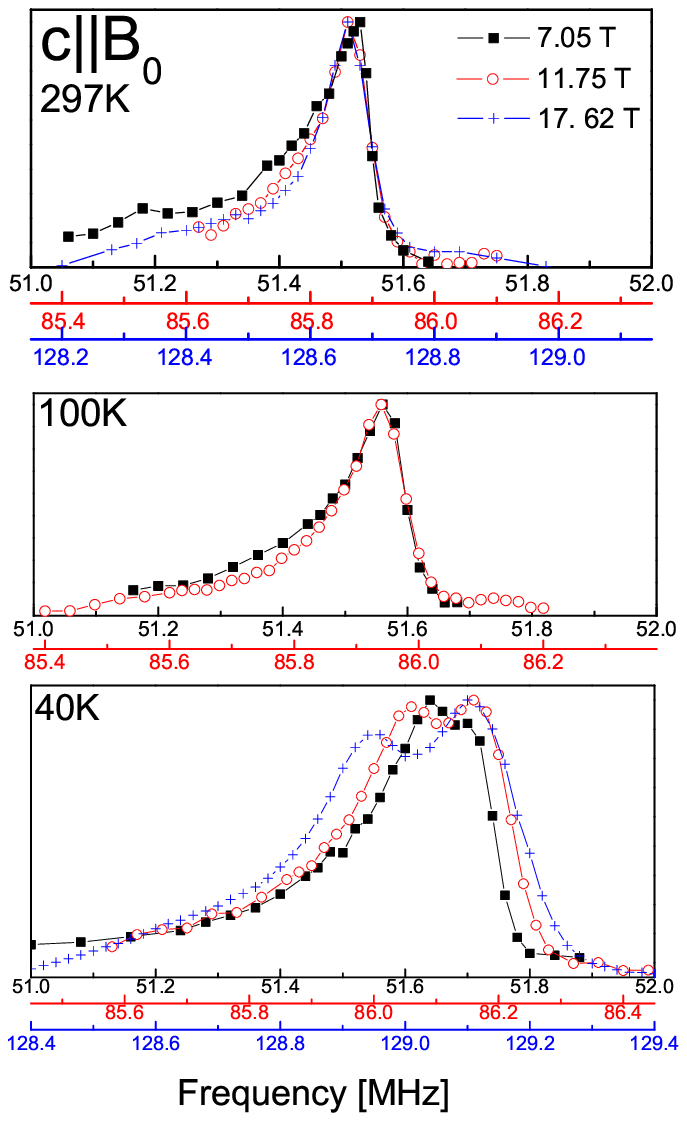}
\caption{Field (7.05 T, 11.75 T and 17.62 T) and temperature (297 K, 100 K and 40 K) dependence of the $^{75}$As spectra for c$\parallel$$B_{0}$. To compare the field dependence of the line shape the frequency range is set to 1 MHz for each spectrum and the spectra are plotted so that the peak frequencies coincide. \label{fig:Fparallel}}
\end{figure}
\begin{figure}
\includegraphics[scale=0.8]{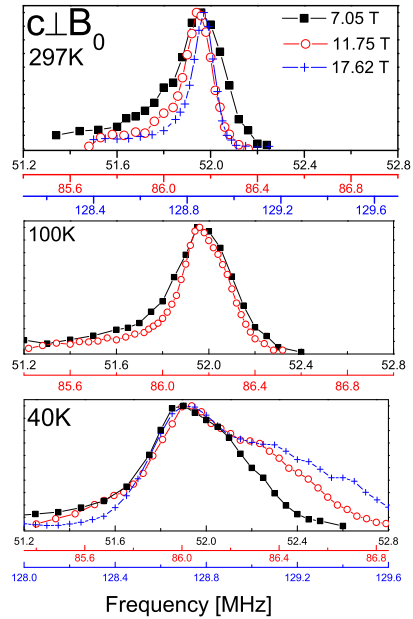}
\caption{Field (7.05 T, 11.75 T and 17.62 T) and temperature (297 K, 100 K and 40 K) dependence of the $^{75}$As spectra for c$\perp$$B_{0}$. To compare the field dependence of the line shape the frequency range is set to 1.6 MHz for each spectrum and the spectra are plotted so that the peak frequencies coincide.\label{fig:Fperp}}
\end{figure}

First we discuss the results for c$\parallel$$B_{0}$ shown in Fig. \ref{fig:Fparallel}. Here the spectra are plotted over the same frequency range and so that the peak frequencies coincide. We want to stress that the NMR shifts are the same at all measured fields and this is consistent with a magnetic NMR shift. We note that for c$\parallel$$B_{0}$ the line has an asymmetric shape with a tail towards lower frequencies (Figs. \ref{fig:Tparallel} and \ref{fig:Fparallel}) while for c$\perp$$B_{0}$ it is more symmetric. However, similar line shapes were observed for aligned LaFeAs$_{0.86}$F$_{0.14}$ \cite{Kitagawa2010}, which suggests that the linewidth asymmetry is intrinsic. The most striking feature of Fig. \ref{fig:Fparallel} is that the linewidth at 297 K and 100 K is almost field independent. This is not expected because in general for c$\parallel$$B_{0}$, the linewidth, $LW_{\parallel}$ comes from a distribution of shifts given in Eq. \ref{eq:shift}. Therefore, for this orientation one expects that $LW_{\parallel}$ is due to a distribution of the spin and orbital shifts, $K_s$ and $K_{L}$ i.e. $LW_{\parallel}$ should be proportional to $B_{0}$. It is likely that there is also a second order quadrupolar contribution to the linewidth, which is $\propto 1/B_0$, and this could also explain the low frequency tail. This could be due to misalignment and/or a variation of the principal axis of the EFG from the $c$ axis. However, the second possibility seems more likely because such asymmetric lines were observed in other well aligned samples.\cite{Kitagawa2010} 

As the temperature decreases there is a significant increase in the linewidth (see Fig. \ref{fig:Tparallel}), which is typical for magnetic broadening.  At 40 K the separation between the two peaks (in kHz) scales linearly with the magnetic field (see Fig. \ref{fig:Fparallel}), which is expected if the peaks arise from coupling to Ce. Additionally, the linewidth of the satellite lines does not change significantly with temperature indicating that the quadrupolar contribution to the width of the central line does not change much with temperature.

Below $T_c$ $LW_{\parallel}$ increases even more (see Fig. \ref{fig:Tparallel}), which is due to further magnetic broadening. At low temperatures there can be additional broadening due to vortex lattice formation in the mixed superconducting state. Vortices induce a distribution of the local magnetic field, $\Delta B_{loc}\propto LW$.\cite{Pincus1964, Brandt1988} However, it would not result in such broad lines as we observe at low temperatures, because typical broadening of NMR lines due to a vortex lattice in pnictides is much smaller.\cite{Matano2008, Ahilan2008, Ma2012}. 

\begin{figure}
\includegraphics[scale=0.6]{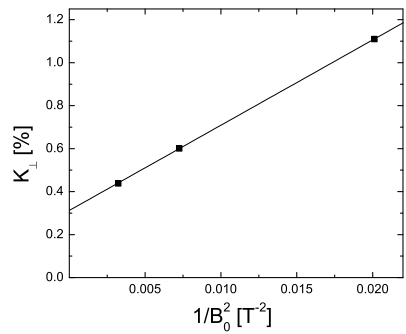}
\caption{Shift $K_{\perp}$ (before subtraction of the second order quadrupolar contribution) versus $1/B^{2}_{0}$ measured at room temperature at fields of 7.05 T, 11.75 T and 17.62 T. A value of $K_{s,\perp}+K_{L,\perp}$=0.31\% for $1/B^{2}_{0}$=0 is obtained from a linear fit to the data points.\label{fig:B2}}
\end{figure}

The field dependence of line shapes for c$\perp$$B_{0}$ is presented in Fig. \ref{fig:Fperp}. For this orientation the total NMR shift contains a sizable second order quadrupolar contribution (Eq. \ref{eq:Kperp}) and it is expected to be proportional to $1/B^{2}_{0}$. We show in Fig. \ref{fig:B2} a plot of the total NMR shift for c$\perp$$B_{0}$ against $1/B^{2}_{0}$ measured at 7.05 T, 11.75 T and 17.62 T and at 297 K (at this temperature there is only a single line in the spectrum). The intercept is $K_{s, \perp} + K_{L, \perp}$ and it amounts to 0.31\%. The gradient is given by $3\nu_{q}^{2}/16$. The value of the second order quadrupolar contribution obtained from a linear fit in Fig. \ref{fig:B2} amounts to 0.29 MHz at 11.75 T. It is slightly bigger than the approximate value of 0.26 MHz estimated from the spectra in Fig. \ref{fig:spectra}. 

For c$\perp$$B_{0}$ the linewidth, $LW_{\perp}$ can be of magnetic and quadrupolar origin, and the latter one should be much bigger compared to c$\parallel$$B_{0}$. At 297 K we observe that the linewidth strongly decreases with increasing field (Fig. \ref{fig:Fperp}) as is expected for quadrupolar broadening. At 100 K the linewidths at 7.05 T, 11.75 T are similar, but this is due to the fact that in the spectrum measured at 11.75 T there is a second peak, which first appears as a high frequency shoulder (see Fig. \ref{fig:Tperp}). With decreasing temperatures $LW_{\perp}$ increases significantly (Fig. \ref{fig:Tperp} and \ref{fig:Fperp}) similar to $LW_{\parallel}$, which arises from line splitting and magnetic broadening induced by the Ce moment that are more noticeable at low temperatures.  

\begin{figure}
\includegraphics[scale=0.75]{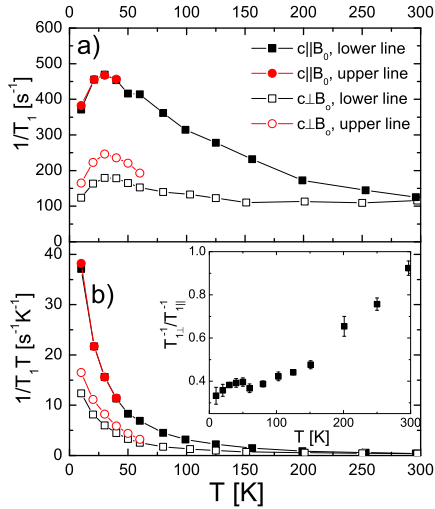}
\caption{a) Temperature dependence of the spin-lattice relaxation rate, $1/T_1$ for c$\parallel$$B_{0}$ and c$\perp$$B_{0}$ measured at 11.75 T.  Lower/upper line refers to the line at lower/higher frequency (see Fig. \ref{fig:Tparallel} and \ref{fig:Tperp}), b) $1/T_1T$ versus temperature. The inset shows the temperature dependence of the $T_{1,\perp}^{-1}/T_{1,\parallel}^{-1}$ ratio calculated for the lower frequency lines. \label{fig:T1}} 
\end{figure}
\begin{figure}
\includegraphics[scale=0.75]{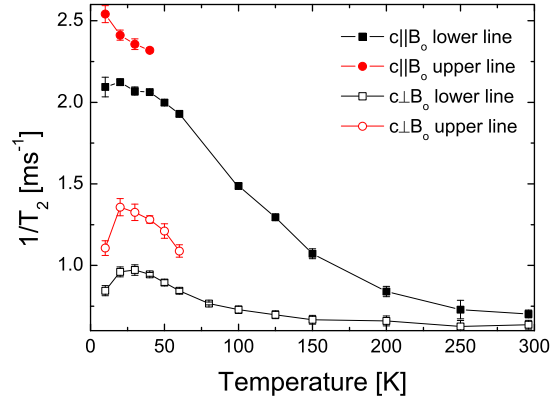}
\caption{Temperature dependence of the spin-spin relaxation rate, $1/T_2$ for c$\parallel$$B_{0}$ and c$\perp$$B_{0}$ measured at 11.75 T. Lower/upper line refers to the line at lower/higher frequency (see Figs. \ref{fig:Tparallel} and \ref{fig:Tperp}) \label{fig:T2}.}
\end{figure}
\subsection{Temperature dependence of $T_1$ and $T_2$}
Equations 1 and 2 were used to calculate the spin-spin and spin-lattice relaxation times and they provided a good fit to the experimental data.  Figures \ref{fig:T1} and \ref{fig:T2} show the temperature dependence of the spin-lattice and spin-spin relaxation rate, $1/T_1$ and $1/T_2$, respectively for c$\parallel$$B_{0}$ and c$\perp$$B_{0}$ orientations. Both relaxation rates increase with decreasing temperature and reach a maximum at a temperature of ~30 K, which is close to $T_c$ and then they start to decrease, except for $1/T_2$ for the upper line for c$\parallel$$B_{0}$, which increases with decreasing temperature. In PrFeAsO$_{0.65}$F$_{0.15}$ and in CeFeAs$_{0.1}$$P_{0.9}$O a similar maximum of $1/T_1$ was observed at $\sim$10 K, which is somewhat lower than in the case of CeFeAsO$_{0.8}$F$_{0.2}$, but still above the magnetic ordering temperature of the rare earth.\cite{Yamashita2010, Sarkar2012}
However, if one plots $1/T_1T$ versus temperature (Fig. 8b) there is no downturn and $1/T_1T$ systematically increases with decreasing temperature down to 10 K.
A strong enhancement of $1/T_1T$ with temperature is markedly different from the behavior observed for LaFeAsO$_{1-x}$F$_{x}$ \cite{Kobayashi2010, Yamashita2010a, Mukuda2009, Mukuda2010} and for the 122 compounds \cite{Ning2010} with similar doping for which it has been found that $1/T_1T$ decreases with decreasing temperature. Our low temperature values of $1/T_1T$ are also more than two orders of magnitude bigger than for LaFeAsO$_{1-x}$F$_{x}$ \cite{Kitagawa2010, Kobayashi2010} or for the 122 pnictides, which must be due to an increase in the spin fluctuations caused by the presence of magnetic Ce ions. 

The anisotropy ratio of the spin lattice relaxation rates defined as $R=\left(T_{1,\perp}^{-1}/T_{1,\parallel}^{-1}\right)$ strongly changes with temperature as can be seen in the inset to Fig. 8b. For the lower frequency lines $R$ is about 0.9 at room temperature and steadily drops to 0.3 at T=10 K, which indicates that the temperature dependence of spin fluctuations is highly anisotropic. $R$ of ~1.2 is observed in LaFeAsO$_{0.86}$F$_{0.14}$ and there is a small decrease to about 1 as the temperature is reduced to 10 K \cite{Kitagawa2010}. The anisotropy ratio of $1/T_2$ also shows a strong temperature dependence. The origin of this temperature dependence in the anisotropy of $1/T_1$ and $1/T_2$ is not clear. However, it suggests that the spin fluctuations probed by $^{75}$As are different when the magnetic field is applied parallel or perpendicular to the c-axis. Since $1/T_1$ is much larger in CeFeAsO$_{0.8}$F$_{0.2}$ than in LaFeAsO$_{1-x}$F$_{x}$ it is possible that this anisotropy is caused by magnetic Ce and strong coupling with Fe spin fluctuations in the FeAs layer.

\subsection{Origin of two $^{75}$As NMR lines} \label{Origin}
First we want to address the interpretation of two $^{75}$As lines given for a very similar compound CeFeAsO$_{0.84}$F$_{0.16}$ \cite{Ghoshray2009}, where two lines were observed only below $T_c$ and attributed to signals from regions inside and outside of the vortices. Their measurements were made at 7.04 T (the smallest field that we used) and only for the c$\perp$$B_{0}$ orientation. If one looks at our Fig. \ref{fig:Fperp} and for an applied magnetic field of 7.05 T it is clear that a strongly asymmetric line shape is observed at 40 K. As noted above, at 11.75 T for c$\perp$$B_{0}$ the second line can already be seen at 100 K as a high frequency shoulder (Fig. \ref{fig:Tperp}). Additionally, the $^{75}$As NMR signals in CeFeAsO$_{0.84}$F$_{0.16}$ were broader, which could cause the second line to become resolvable in the NMR spectrum only at lower temperatures when the line splitting becomes large. Therefore, we believe that the fact that in CeFeAsO$_{0.84}$F$_{0.16}$ \cite{Ghoshray2009} two lines were observed only below $T_c$ could be accidental and due to additional line broadening because of incomplete alignment and hence the two peaks are not due to the effect of vortices in the superconducting state.       

As was mentioned above in the Experimental section there is only a very small amount of impurities containing As in the sample, i.e. FeAs$_2$ and FeAs. Neither of them magnetically orders around 100-120 K and hence they can not explain the shift behavior or appearance of the second line. FeAs orders below 77 K \cite{Selte72, Yuzuri80} and FeAs$_2$ does not order at all \cite{Yuzuri80}.
We also did not observe any sudden changes of the $Q$-factor of the probehead, which would be the case if a significant amount of sample magnetically ordered.

One can try to estimate the relative signal intensity of both lines. It can be done most reliably for c$\parallel$$B_{0}$ at 17.62 T and at 40 K, i.e. still above $T_c$ to avoid possible signal loss of one or two lines due to the superconducting state. For this measurement the two lines overlap the least and the estimated intensity ratio amounts to about 1:2. Therefore, none of the lines can come from the impurity phase. Additionally, the integrated NMR signal intensity corrected for temperature, $Q$-factor and $T_2$ did not change when the second line appeared, which is consistent with the NMR signal arising from the same phase fraction for all of the measured temperatures.

At a given temperature two lines are observed at different frequencies for different orientation (compare Fig. \ref{fig:Tparallel} and \ref{fig:Tperp}). This excludes the possibility that we observe two lines due to poor sample alignment. Similar temperature behavior of the shifts, $1/T_1$ and $1/T_2$ indicates that the signals come from $^{75}$As nuclei in regions that have similar electronic properties. 

It is possible that the two lines arise from ordered and disordered regions in the FeAs layer. In this scenario the temperature-dependent $^{75}$As NMR shift arises from hyperfine coupling from Ce to the As. In the ordered region it is possible that there is superexchange between the Ce moments via the FeAs conducting layer (possibly via an RKKY interaction), as has already been suggested from $^{17}$F NMR measurements on SmFeAsO$_{1-x}$F$_x$\cite{Prando2010}. This leads to a non-zero Curie-Weiss temperature, which is negative in our case and consistent with Ce antiferromagnetic order that occurs at low temperature, low doping, and via an exchange interaction with Fe. In the disordered region it is possible that the average Ce-Ce superexchange energy is zero, which could lead to the observed Curie-Weiss temperature that is near zero (e.g. by disrupting the RKKY interaction). The question is then why are there ordered and disordered regions in the FeAs layer? It may be related to charge disorder that occurs near the F dopant ion or possibly due to an inhomogeneous F concentration and/or oxygen vacancies in the CeO$_{0.8}$F$_{0.2}$ layer. The differences in the Curie-Weiss temperature from \mbox{-18 K} in the ordered to ~0 K in the disordered region is not very large and hence it is possible that only a small amount of disorder is sufficient to reduce the Ce-Ce superexchange energy. As we have already noted, disorder induced affects have been observed in other ReFeAsO$_{1-x}$F$_x$ (Re is a rare earth) NMR measurements either by the appearance of additional low temperature NMR peaks or two different spin-lattice relaxation rates.

\section{Conclusions}
In conclusion, $^{75}$As NMR measurements on CeFeAsO$_{0.8}$F$_{0.2}$ show the appearance of two peaks in the $^{75}$As NMR spectra that are evident for temperatures as high as 100 K, which is well above the superconducting transition temperature of 39 K. Thus, contrary to a previous $^{75}$As NMR study at a lower magnetic field, on less well aligned samples and for lower doping, the 2 peaks cannot be attributed to the effect of vortices. It is likely that the 2 peaks arise from hyperfine coupling from magnetic Ce to As and ordered and disordered regions in the CeO layer where the disorder could be due to regions where the F concentration is highly inhomogeneous and/or regions with clusters of oxygen vacancies. We find that this interpretation is consistent with the temperature dependence of the $^{75}$As NMR shift data where the 2 peaks can be fitted to Curie-Weiss-like terms with a Curie-Weiss temperature of near zero or $\sim$-18 K. Interestingly we find that the hyperfine coupling from Ce to As changes sign when changing the applied field from parallel to perpendicular to the $c$-axis and this ratio is $\sim$-1.5. The $^{75}$As spin-lattice relaxation rates are dominated by coupling of Ce moment spin fluctuations to $^{75}$As in the low temperature region where 1/$T_1T$ systematically increases with decreasing temperature. The ratio of the spin-lattice relaxation rate, R, is near 1 at room temperature and decreases with decreasing temperature, which could be due to coupling between the Ce moment and the FeAs layer that is stronger when the magnetic field is applied along the $c$-axis direction. Similar to some studies on other 1111 compounds we find two $T_1$ and two $T_2$ values at low temperatures (above T$_c$). This is not surprising when there are ordered and disordered regions and suggests that disordered regions in the ReO layer may be a common feature in the 1111 compounds. Since doping in the FeAs layer is via charge transfer from the insulating ReO layer, then disordered ReO regions should lead to electronic disorder in the conducting FeAs layer.

\section{Acknowledgments}
We acknowledge financial support from the University of Leipzig, IRTG: Diffusion in porous media (T.M.), the New Zealand Marsden Fund (VUW0917), the MacDiarmid Institute for Advanced Materials and Nanotechnology, and the assistance of M. Jurkutat and J. Kohlrautz (Leipzig). 

\bibliography{Pnictide}

\end{document}